# Intelligent Breast Cancer Diagnosis with Heuristic-assisted Trans-Res-U-Net and Multiscale DenseNet using Mammogram Images

Muhammad Yaqub[1], Feng Jinchao[1]
[1]Faculty of Information Technology, Beijing University of Technology, Beijing, China.

**Abstract-** Breast cancer (BC) significantly contributes to cancer-related mortality in women, underscoring the criticality of early detection for optimal patient outcomes. A mammography is a key tool for identifying and diagnosing breast abnormalities; however, accurately distinguishing malignant mass lesions remains challenging. To address this issue, we propose a novel deep learning approach for BC screening utilizing mammography images. Our proposed model comprises three distinct stages: data collection from established benchmark sources, image segmentation employing an Atrous Convolution-based Attentive and Adaptive Trans-Res-UNet (ACA-ATRUNet) architecture, and BC identification via an Atrous Convolution-based Attentive and Adaptive Multi-scale DenseNet (ACA-AMDN) model. The hyperparameters within the ACA-ATRUNet and ACA-AMDN models are optimised using the Modified Mussel Length-based Eurasian Oystercatcher Optimization (MML-EOO) algorithm. Performance evaluation, leveraging multiple metrics, is conducted, and a comparative analysis against conventional methods is presented. Our experimental findings reveal that the proposed BC detection framework attains superior precision rates in early disease detection, demonstrating its potential to enhance mammography-based screening methodologies.

## 1. Introduction

The most prevalent type of malignancy in women is BC. Next to cancer, it is the second leading reason of mortality in women [1]. One in every 36 female deaths is related to BC, or around 3% of all female deaths are caused by BC. In order to improve the survival rate of the patient, early BC identification is crucial [2]. Researchers are introducing increasingly accurate models for BC diagnosis into practice because of the tremendous fatality and high expense of cancer-related treatment [3, 4]. Radiotherapists use mammography as an efficient imaging method to detect and screen the presence of BC. Mammography is the primary clinical test for BC and is quite accurate in predicting BC. Breast Mammography usually detects 80–90% of BC cases in females with no symptoms [5]. Breast lumps and calcifications are considered the early signs of BC, respectively. The most effective method for detecting BC is mammography [6].

BC is identified as early as possible because of mammography. As a result, the cost of treating BC is reduced drastically. Although mammography is advantageous, screening BC from it consumes more time and requires expert knowledge for accurate detection. Also, the decision made by radiologists varies. Computer Assisted Diagnosis (CAD) software is created to help radiotherapists improve prediction accuracy while screening through mammography [7]. But CAD software has a substantial risk of producing false positive and false negative outcomes [8]. As a result, it is crucial to improve doctors' detection effectiveness using CAD systems with deep learning approaches. Because of their diverse size, shape, and position, the BC cells are difficult to detect and localise automatically from the BC mammography images [9].

Several methods have been developed for fragmenting medical mammography images. Due to the limitations of each approach, segmentation of mammography images is proven to be a challenging task [10]. The recognition of signs of BC from breast mammography images is also a significant challenge. An essential stage in the analysis of mammography images is the extraction mechanism. Traditional approaches use handcrafted features to describe the image's content [11]. The neural network is developed as a replacement for automatically detecting the best characteristics. Additionally, incorrectly interpreting these images results in a dangerously inaccurate diagnosis [12]. Consider a false negative diagnosis, where a BC in its early stages is mistaken for a typical instance. One of the traditional methods for reducing these kinds of computational errors is feature selection (FS), which eliminates duplicate features and chooses a group of distinctive features. Distinctive features might not be granted the significance they should require in the classification process due to these duplicate features [13].

To aid in decision-making, experts have used a variety of machine learning [14] approaches in medical image interpretation over the last few years. But the system's performance suffers concerning efficacy and precision because of the complicated nature of traditional machine learning methods, like segmentation, feature extraction, preprocessing, and others [15, 16]. The newly developed deep learning techniques tackle conventional machine learning problems. This technique can successfully represent features to perform the tasks of object localisation and image classification [17]. Medical professionals can use their expertise to connect dataset features to facts, which is difficult for machine learning techniques to accomplish. In contrast to conventional techniques that rely on manual methods, deep learning eliminates this problem by including processing and future engineering as a component of learning [18]. CNN is the most widely used deep learning method for image processing. The 2D input-image configuration can specifically alter the CNN design. However, it achieved a precision of only 88%, which has to be further raised to be more effective than the state-of-the-art methods [19]. Hence, this paper generates and implements a novel deep-learning approach for early screening and BC.

The main achievements that have been performed in this work are listed below.
- To build a highly evolved deep learning-assisted BC detection framework with a heuristic algorithm to increase efficiency and precision and reduce the error in BC detection.
- To implement a newly developed advanced deep learning model called ACA-ATRUNet for precisely segmenting the BC mammogram images, which constitute the combination of Transformer, ResNet, and UNet with attention mechanism with parameters optimisation by MML-EOO algorithm.
- To detect the BC from segmented mammogram images with the help of a newly implemented deep learning framework called ACA-AMDN along with parameters tuning with the help of MML-EOO algorithm for accurate detection of BC and minimising the false rate during the detection stage.
- To develop a highly efficient heuristic algorithm called MML-EO algorithm for tuning the parameters like hidden neurons and epochs in both ACA-ATRUNet, and ACA-AMDN classifiers, batch size in ACA-AMDN classifiers, and steps per epochs in ACA-ATRUNet classifier, respectively to enhance the detection accuracy.
- To validate the performance of the proposed model with other classifiers and algorithms to prove the better performance of the proposed model in BC detection.

The further sections in this paper are given below. Section II carries the literature review done in examining the pre-existing works. Section III contains the development of the intelligent BC detection model with heuristic-assisted Trans-Res-UNet and Multiscale DenseNet using mammogram images. Section IV provides a detailed view of the Atrous Convolution-based Attentive and Adaptive BC segmentation model using mammogram images. Section V comprises the architectural representation of the Atrous Convolution-based Attentive and Adaptive BC detection model using mammogram images. Section VI gives the experimental outputs and the discussions that are carried out regarding the generated results. Section VII summarizes the developed deep learning-based BC detection framework. Section VIII contains the supplementary material, Appendix A, which provides additional details and supporting information for this study.

## 2. Motivation

### 2.1 Literature review

Das et al. (2021) proposed a stacked ensemble model for breast cancer (BC) classification that combined breast histopathology images and gene expression data [20]. The model incorporated the Convex Hull Algorithm and t-Distributed Stochastic Neighbor embedding to transform the 1D gene expression data into images. The dataset and its decomposed forms were utilised to enhance performance. Three convolutional neural networks (CNNs) served as the foundational classifiers in the first stage, with the data decomposed using Variational Mode Decomposition and Empirical Wavelet Transform. The results of the first stage were used to train the second stage classifier. The model employed gene expression data from Mendeley to create 2D datasets. Training and validation were conducted using synthetic and photographic datasets of breast histopathology. The proposed method demonstrated improved performance, highlighting its effectiveness in BC classification.

In 2021, Saber et al. [21] proposed a Transfer Learning (TL) method using ResNet50, Inception V3, VGG-16, ResNet, and VGG-19 networks for feature extraction from the MIAS dataset to aid in the automatic identification and classification of breast cancer (BC) susceptible areas. TL of the VGG16 model demonstrated effective categorization of mammogram breast images for BC diagnosis. In 2022, Jiang et al. [22] introduced the Probabilistic Anchor Assignment (PAA) technique to accurately identify and classify mammograms as malignant or benign, improving prognosis ability. The proposed framework included a single-stage PAA-based detector to identify abnormal tumor areas and a two-branch ROI detector for tumor categorization, incorporating a Threshold-adaptive Post-processing Algorithm for complex breast data. The model was trained and evaluated using publicly available mammogram databases, demonstrating enhanced classification accuracy compared to existing techniques.

In 2022, Kavitha et al. [23] presented a novel computerized mammogram-based breast cancer (BC) diagnosis framework. The framework employed median filtering for preprocessing to eliminate irrelevant data in mammographic images. BC segmentation was achieved using the Optimal Kapur's Multilevel Thresholding with Shell Game Optimization (OKMT-SGO) method. The proposed model incorporated a Backpropagation Neural Network (BPNN) classifier and a CapsNet feature extractor for BC identification. Evaluation using benchmark DDSM and Mini-MIAS datasets showcased the superior performance of the proposed method in diagnostic accuracy. In 2022, Kumari and Jagadesh [24] employed feature selection techniques to enhance classifier performance. Intensity, texture, and shape-based features were extracted from preprocessed medical images. The selected features were used with the XGBoost classifier and compared to other classifiers. The MIAS database was utilized for experimentation and evaluation. Results showed that the proposed XGBoost framework outperformed other feature selection techniques in categorising MIAS mammography images as abnormal or normal, demonstrating its superior performance.

In 2021, Patil and Biradar [25] proposed an improved hybrid classification model for mammogram-based breast cancer (BC) detection. The approach involved image preprocessing, feature extraction, segmentation [26, 27], and identification stages. A median filter was used for noise removal, and the Firefly updated Chicken-based Optimization (FC-CSO) algorithm was employed for tumor segmentation. Features were extracted and fed into a Recurrent Neural Network (RNN) and a Convolutional Neural Network (CNN) for classification [28]. The combination of the two models achieved superior accuracy compared to traditional classifiers.

In 2022, Pramanik et al. [29] proposed a breast mass categorization system using mammograms. The VGG16 architecture with an attention mechanism extracted deep features from mammography images. The Social Ski-Driver (SSD) technique and Adaptive Beta Hill Climbing search strategy were employed to obtain optimal features. The K-Nearest Neighbors (KNN) classifier utilized these features for data classification. The suggested model demonstrated successful recognition and discrimination between healthy and cancerous breasts. Remarkably, the framework achieved higher precision by utilizing only 25% of the attention aided VGG16 model's features on a publicly available dataset. In 2020, Zheng et al. [30] introduced the DL-Assisted Efficient AdaBoost Algorithm (DLA-EABA) for breast cancer (BC) identification. The study investigated the characterization of breast masses using transfer learning from diverse imaging modalities, such as mammography, MRI, digital breast tomosynthesis, and ultrasound. The deep learning model incorporated LSTM layers, convolutional layers, fully connected layers, max-pooling layers, activation layers, and error estimation for classification. The fusion of machine learning approaches with feature selection and extraction methods was examined, and the model's performance was evaluated against existing segmentation methods and conventional classifiers.

## 2.2 Problem Statement and Objectives

Breast cancer screening through mammography is crucial in early detection, offering the potential for more successful treatment outcomes. However, challenges arise in accurately interpreting mammograms, particularly in women with dense breasts, leading to increased rates of false predictions. Furthermore, normal mammogram results do not guarantee the absence of breast cancer, underscoring the limitations of relying solely on this screening method. In addition, false diagnoses can subject women to unnecessary radiation exposure. The complexity of handling the large volume of mammography images and the variability in prediction outcomes among radiologists further highlight the limitations of traditional breast cancer detection approaches. To address these challenges and improve the accuracy of breast cancer prediction while minimizing errors, we propose adopting deep learning techniques for breast cancer detection from mammography images. This novel approach aims to enhance the precision of predictions and mitigate the generation of false errors, offering a more efficient and reliable method for breast cancer detection.

Table 1 presents a comprehensive overview of current breast cancer (BC) detection techniques, along with their respective merits and demerits [20-25, 29, 30]. CNN and EWT demonstrate improved accuracy, recall, and precision detection rates but introduce hardware and time complexity. VGG achieves higher accuracy, AUC, and sensitivity, enhancing system robustness, primarily for prognosis objectives. PAA extracts peripheral regions to identify diseases but incurs computational burdens that degrade system robustness. BPNN offers discriminative features for detection but lacks parameter tuning for further enhancement. XGBoost selects notable features for increased detection accuracy but does not support large-scale dimensional datasets. CNN significantly extracts boundary-level regions for accurate results yet blur or noise in images degrades system robustness and can lead to misdiagnosis. KNN acquires deep and optimal features for cancer region detection but suffers from high time complexity and premature convergence. CNN-LSTM obtains desirable early-stage disease detection but is susceptible to overfitting. These challenges underscore the need to develop and implement an accurate BC detection approach using deep learning techniques.

**Table 1.** Merits and demerits of traditional BC detection methods.

| Author [citation] | Methodology | Features | Challenges |
|---|---|---|---|
| Das et al. [20] | CNN and EWT | • It improves the detection rate regarding the accuracy, recall, precision, etc. | • It causes hardware complexity as well as time complexity. |
| Saber et al. [21] | VGG | • It achieves more accuracy, AUC, and sensitivity, enhancing the system's robustness. | • It is further developing for prognosis objectives. |
| Jiang et al. [22] | PAA | • It extracts the peripheral regions to get the features for identifying the diseases. | • It causes a computational burden that degrades the robustness of the system. |
| Kavitha et al. [23] | BPNN | • It offers discriminative features for reaching a more detection rate. | • It does not tune the parameters used in the model for further enhancement. |
| Kumari and Jagadesh [24] | XGBoost | • It chooses the noteworthy features for increasing the detection accuracy. | • It does not support large-scale dimensional datasets. |
| Patil and Biradar [25] | CNN | • It significantly extracts the boundary-level regions for estimating the appropriate results. | • The blur or noise present in images degrade the system's robustness and misdiagnoses the disease. |
| Pramanik et al. [29] | KNN | • It acquires deep and optimal features for detecting the cancer regions in images. | • Time complexity and premature convergence rate occur. |
| Zheng et al. [30] | CNN-LSTM | • It obtains the desired value to detect the disease at its early stages. | • It causes the overfitting problem. |

## 3. Intelligent Breast Cancer Detection Model with Heuristic-assisted Trans-Res-U-Net and Multiscale DenseNet using Mammogram Images

### 3.1 Proposed Model and Description

BC is the major type of cancer that is commonly found in women. The early detection of BC can save many lives. Unfortunately, the early detection of BC is rare because of the drawbacks in the available BC detection techniques. The image processing technique is most commonly used in the early diagnosis of BC. Some frequently used image processing techniques include breast ultrasound, MRI, CAD tools, and mammography. With the help of these image-processing techniques, radiologists can diagnose the presence of BC in women at a very early stage. Prognosis is the only way to detect BC at an early stage. However, the results obtained from the image processing techniques rely on the perception and knowledge of the radiologist, which may result in unwanted misdiagnosis and wrong prediction results. Also, these image-processing techniques do not give highly accurate and precise results for women with dense breast tissues. Women with surgical histories also suffer from misdiagnosis. Due to misdiagnosis, healthy women may also undergo unwanted radiation exposure, which results in further health effects. And also, these image-processing techniques are expensive. Further detection techniques, such as Positron Emission Tomography (PET), Electrical Impedance Imaging, Optical Imaging Tests, Molecular Breast Imaging, Nuclear Medicine Studies, and Galactograms, have been developed to overcome these issues. However, these techniques are costly and can detect BC in women with higher risk rates alone. A newly evolved deep-leaning-based BC detection framework is generated to overcome these issues.

In the developed deep learning-based BC detection framework, the mammogram images are primarily obtained from standardised mammogram image data sources. These raw images are then provided to the developed ACA-ATRUNet classifier for the segmentation process. Before classification, these images are segmented to improve the overall accuracy of the further detection process. The segmented images are now given to the implemented ACA-AMDN framework for classifying the BC images. An enhanced metaheuristic optimisation algorithm called the MML-EOO algorithm is suggested to reduce the processing complexity and computational time. The recommended MML-EOO algorithm optimises the hidden neurons in the ACA-ATRUNet classifier, Epochs in the ACA-ATRUNet classifier, steps per epochs in the ACA-ATRUNet classifier, hidden neurons in the ACA-AMDN classifier, epochs in the ACA-AMDN classifier, and the batch size in the ACA-AMDN classifier, respectively. These optimised parameters help in fastening the entire detection process. The final detection image of BC is obtained from the ACA-AMDN classifier.

### 3.2 Mammogram Image Collection

Two major BC mammography image databases provided the input BC pictures needed to carry out the segmentation and detection functions in the implemented BC detection model. Table 2 contains information about the database and the sources from which the images are available. The term $BC_{fs}^{img}$ represents the collected images from the two standard databases.

Table 2. Description of a mammography image database.

| Sr. No. | Database Name | Image Description | Available Website |
|---|---|---|---|
| 1 | MIAS Mammography | Images and labels for mammography scans make up the database. This dataset consists of seven labels: reference number, background tissue's characteristics, abnormality class, condition of abnormality, image coordinates, and radius of abnormality. The images provided in this database are of size 1024 x 1024 pixels. | "https://www.kaggle.com/datasets/kmader/mias-mammography access date: 2023-01-11" |
| 2 | CBIS-DDSM: Breast Cancer Image Dataset | There are 2,620 digitised film mammography studies in the DDSM database. It includes instances with certified pathology data for normal, benign, and cancerous conditions. The DDSM is a helpful tool in designing and testing decision support systems due to the size of the database and ground truth checking. A trained radiologist chose and curated a subset of the DDDSM data for the CBIS-DDSM collection. After being decompressed, the images were changed to DICOM format. Together with pathologic diagnosis for training data, updated ROI segmentation and bounding boxes are also given. | "https://www.kaggle.com/datasets/awsaf49/cbis-ddsm-breast-cancer-image-dataset access date: 2023-03-07" |

The images are gathered from the two databases mentioned above, shown in Figure. 1.

| Image Type/Number | Sample Image 1 | Sample Image 2 | Sample Image 3 | Sample Image 4 | Sample Image 5 |
|---|---|---|---|---|---|
| MIAS Mammography Dataset | | | | | |
| Abnormal | | | | | |
| Normal | | | | | |
| CBIS-DDSM Breast Cancer Image Dataset | | | | | |
| Benign without callback | | | | | |
| Benign | | | | | |

| Malignant | 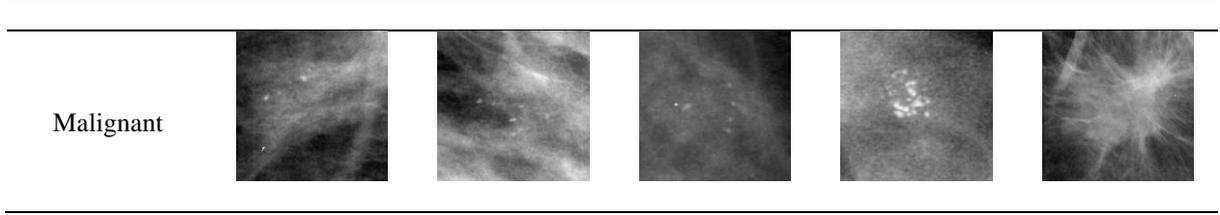 |
|---|---|

**Figure 1.** Sample Images Gathered from the MIAS and CBIS-DDSM Database

## 4. Atrous Convolution-based Attentive and Adaptive Breast Cancer Segmentation Model using Mammogram Images

### 4.1 ACA-ATRUNet-based Breast Cancer Segmentation

The obtained raw mammogram images $BC_{fs}^{img}$ are given to the generated ACA-ATRUNet for image segmentation. ACA-ATRUNet is developed by replacing the normal convolutional layer in the Trans-Res-UNet with an Atrous convolutional layer and including an attention mechanism. The entire process is repeated several numbers of times (Multiscale) before producing the final segmented BC image output $SI_{ad}^{TRU}$. With the aid of the recommended MML-EOO algorithm, the elements in the ACA-ATRUNet structure are adjusted to increase the precision of the segmented BC images. The elements, such as epochs, steps per epochs, and the hidden neurons in the ACA-ATRUNet, are optimized by the suggested MML-EOO algorithm. This optimization is done to maximize the dice coefficient and the accuracy between the mask images and the segmented BC images. The main idea behind this parameter optimization is mathematically represented in Eq. (1).

$$S1 = \underset{\{hi_{lm}^{TRU}, eh_{kl}^{TRU}, se_{jk}^{TRU}\}}{\arg\min} \left( \frac{1}{Dice + Arcy} \right) \quad (1)$$

The term $S1$ in Eq. (4) denotes the objective function of the developed ACA-ATRUNet, $hi_{lm}^{TRU}$ denotes the optimally adjusted number of hidden neurons, $eh_{kl}^{TRU}$ denotes the optimally adjusted number of epochs, $se_{jk}^{TRU}$ denotes the number of optimally adjusted steps per epoch, $Dice$ denotes the dice co-efficient between the mask image and the segmented BC image, and $Arcy$ denotes the accuracy. The steps per epoch are tuned in the range [300,1000], the hidden neurons are tuned in the range [5,255], and the epochs are tuned in the range [5,50]. These parameters are tuned to maximise the Dice coefficient and accuracy. The Dice coefficient is the overlap among the masked and segmented images. The dice coefficient between the mask image and the segmented image is given by Eq. (2).

$$Dice(ML_{ma}^{am}, SI_{ad}^{TRU}) = \frac{2(MI_{ma}^{am} \cap SI_{ad}^{TRU})}{MI_{ma}^{am} + SI_{ad}^{TRU}} \quad (2)$$

The term $MI_{ma}^{am}$ in Eq. (5) denotes the mask images and $SI_{ad}^{TRU}$ denotes the segmented BC image. The accuracy $Arcy$ evaluated using Eq. (3).

$$Arcy = \frac{TU + VW}{TU + TV + VW + VX} \quad (3)$$

In Eq. (6), the terms $VW$ represent the true negative, $TU$ represents the true positive, $VX$ represents the false negative, and $TV$ represents the false positive, respectively. The pictorial representation of the implemented ACA-ATRUNEt-based BC mammogram image segmentation is provided in Figure. 2.

**Figure 2.** Pictorial representation of implemented ACA-ATRUNet-based BC mammogram image segmentation.

## 5. Architectural Representation of Atrous Convolution-based Attentive and Adaptive Breast Cancer Detection Model using Mammogram Images

### 5.1 ACA-AMDN-based Breast Cancer Detection

The segmented images from the ACA-ATRUNet $SI_{ad}^{TRU}$ are fed to the developed ACA-AMDN structure for BC image classification. ACA-AMDN is developed by replacing the normal convolutional layer in the DenseNet with an Atrous convolutional layer and including an attention mechanism. The process is repeated several times (Multiscale) in the DenseNet structure before producing the final classification output. The classified image output of BC mammogram images is given by $CI_{hs}^{MDN}$. The parameters are optimized in the ACA-AMDN structure with the assistance of the implemented MML-EOO algorithm. The parameters like epochs, batch size, and the hidden neurons in the Multi-scale DenseNet are optimally tuned with the help of the proposed MML-EOO algorithm. This optimization aims at maximizing accuracy and minimizing False Positive Rate (FPR). The major contribution behind this parameter optimization is formulated as in Eq. (4)

$$S2 = \underset{\{hi_{ml}^{MDN}, eh_{lk}^{MDN}, bs_{kj}^{MDN}\}}{\arg\min} \left( \frac{1}{Arcy} + XY \right) \tag{4}$$

The term $S2$ in Eq. (8) denotes the objective function of the developed ACA-AMDN, $hi_{ml}^{MDN}$ denotes the optimally adjusted number of hidden neurons, $eh_{lk}^{MDN}$ denotes the optimally adjusted number of epochs, $bs_{kj}^{MGN}$ denotes the number of optimally adjusted steps per epoch, and $XY$ denotes the FPR. The batch size is tuned as [2,4,8,16,32,64] the hidden neurons are tuned in the range [5,255], and the epochs are tuned in the range [5,50]. These parameters are optimized to maximize the accuracy and minimize the FPR. The FPR is computed using Eq. (5) as follows.

$$XY = \frac{TV}{VX + TV} \tag{5}$$

The pictorial illustration of the implemented ACA-AMDN-based BC classification is shown in Figure. 3.

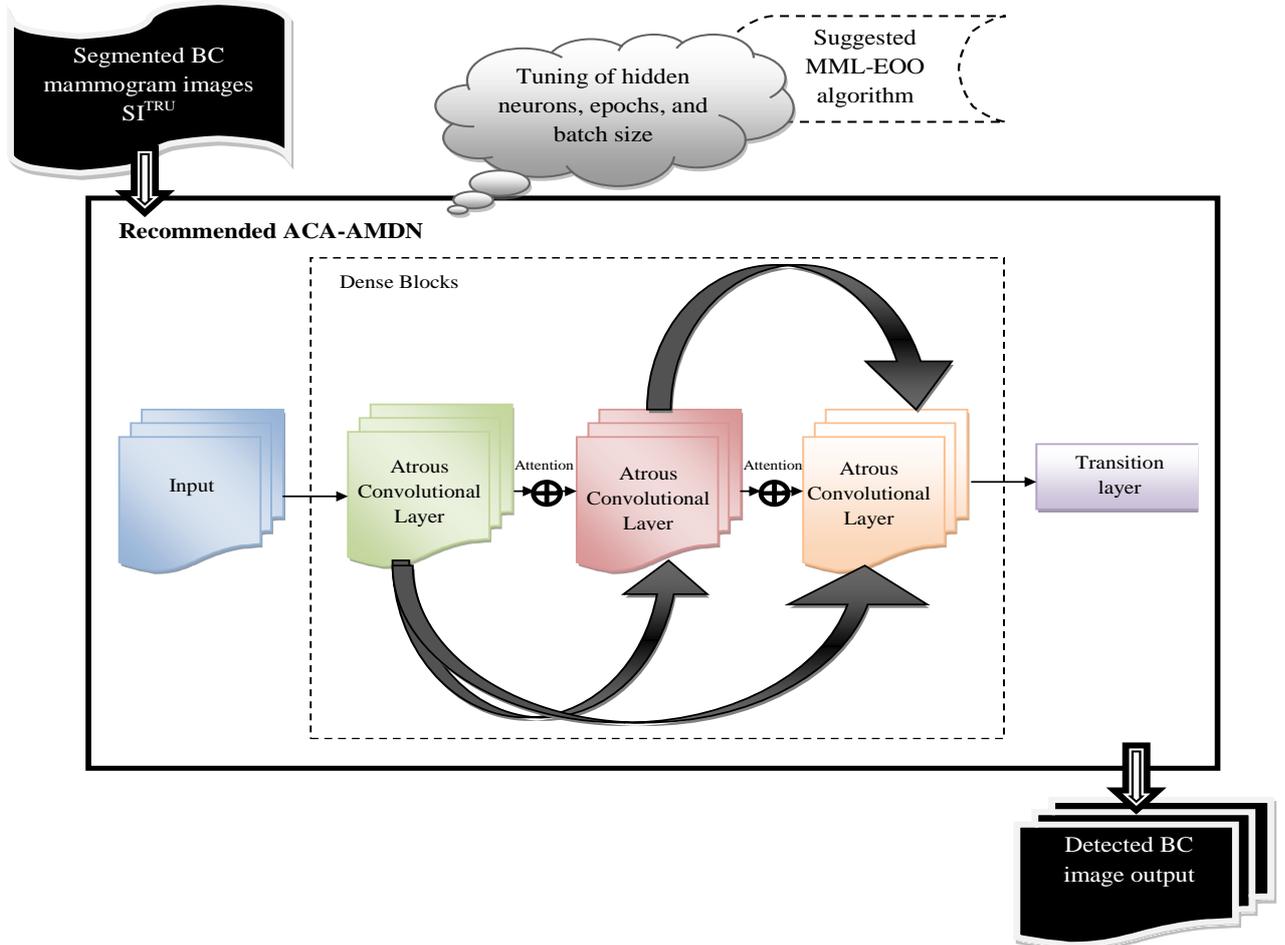

**Figure 3.** A pictorial illustration of the implemented ACA-AMDN-based BC classification model.

## 5.2 Proposed MML-EOO

By optimizing Trans-Rs-UNet components like epochs, hidden neurons, and the steps in the epochs as well as Multiscale DenseNet characteristics like epochs, hidden neurons, and batch size, the generated BC classification model's final prediction result can be improved. The suggested MML-EOO algorithm achieves this parameter optimization. Because of its balanced exploitation and exploration, and the capacity to eliminate local optimums, the EOO [30] algorithm is used in this paper. Due to the oyster size selection constraint, however, this technique cannot resolve challenging real-time issues. As a result, the EOO algorithm's oyster size constraint $O$ is upgraded using the formula provided in Eq. (6).

$$O = 5 - r * \left(\frac{2}{R}\right) \tag{6}$$

The term $r$ in Eq. (10) denotes the current iteration value, $R$ denotes the maximum iteration count, and $O$ denotes the size of the oyster. The value of $O$ is in the range $[3,5]$ in the traditional EOO algorithm, which is upgraded using Eq. (10) in the developed MML-EOO algorithm. The value $O$ decreases linearly from 50mm to 30mm in the suggested MML-EOO algorithm. The value $O$ in Eq. (6) is used to update the size of the oyster in Eq. (7), Eq. (8), Eq. (9), and Eq. (10).

**EOO [31]:** The EOO algorithm was driven by the food choices of the bird Eurasian Oystercatchers (EO) as they searched for appropriate oysters. EOO was invented by monitoring the behavior of EO during oyster hunting and also how they consume the oyster that has been caught. To make up for the energy it loses when breaking the shell of the oyster, the EO must concentrate on eating an enormous volume of the oyster. In real-world situations, the bird doesn't pick huge oysters. Applying two assumptions allows for the resolution of this paradox. 1) The first presumption is that since certain oysters have extraordinarily hard shells that cannot be cracked, the average benefit from huge oysters falls below the maximum advantage. Sometimes EO selects huge oysters that, despite their best efforts, cannot break. The time invested in handling the huge, unbreakable oysters diminishes the average benefit derived from these oysters. When this aspect is considered, it is expected that the EO should concentrate on oysters around 30 and 45 mm in size. 2) The second theory holds that the huge oysters' shells are covered in a layer of barnacles. As a result, it is nearly impossible to open the shell of that oyster, and the EO does not like to hunt this kind of oyster. Even though they have more calories, this bird does not choose oysters which barnacles coat. Larger oysters may have more barnacles coating them, making them potentially indestructible and not a recommended option. Suppose the effort required to open the selected oysters is considered in the mathematical model, together with the time lost trying to open certain unbreakable oysters and the appropriate sizes for the attack. In that case, the EO must concentrate on oysters between 30 and 45 mm. So, in the EOO algorithm, obtaining the ideal balance between caloric intake, wasted time, and energy gained is crucial to obtaining the best solution. The main objective of the EOO algorithm is to balance the wasted calories in opening the oyster and the energy obtained from the oyster. The amount of energy required to crack the oyster's shell, the oyster's size, and the number of calories in the oyster's flesh are all strongly connected. The size of the oyster affects both the number of calories ingested by the EO and the time the EO takes to open the oyster. As a result, EO uses a lot of energy to extract the oyster from its shell. The exploration of the EO is described as follows. The amount of energy that is available in the EO $K$ at the final stage of hunting the oyster is given by Eq. (7).

$$K = J + N + O * f * (P_m - P_{r-1}) \tag{7}$$

This size of the oyster $O$ in Eq. (7) is upgraded using the fitness-based concept provided in Eq. (6). In Eq. (11) $N$ denotes the current energy requirement, $J$ denotes the time requirement of the EO to open the ideal oyster, and $f$ represents a number in random in the range $[0,1]$ that is selected to increase the predictability in the search area. The value of the available energy in the EO $K$ varies inversely as the iteration count $r$. The position in which the ideal oyster is found available is provided in Eq. (8).

$$P_r = P_{r-1} * Q \tag{8}$$

The term $Q$ in Eq. (8) represents the amount of energy the EO obtained from eating the ideal oyster of size $O$ and $P_r$ represents the position of the ideal oyster. The value of $J$ and the value of $Q$ relies on $O$. The time required to open a selected oyster $J$ is formulated as in Eq. (5).

$$J = \left(\left(\frac{O-3}{5-3}\right) * 10\right) - 5 \tag{9}$$

The value of $O$ in Eq. (9) is updated using the fitness-based concept provided in Eq. (10). The presently available energy in the bird is computed as in Eq. (10).

$$D = \left(\frac{r-1}{s-1}\right) - 0.5; r > 1 \tag{10}$$

The calorie that can be obtained by consuming the oyster $Q$ is given in Eq. (11).

$$Q = \left(\left(\frac{O-3}{5-3}\right)*2\right) + 0.6 \tag{11}$$

The value of $Q$ in Eq. (11) is updated using Eq. (6). If the time is negative, it represents that the bird has reached its maximum capacity in opening the oyster and cannot further spend energy in opening it. This is considered an exceptional case. $N$ remains constant in the last iteration and its preceding iteration. Thus $N$ and $J$ will have a negative value. The main contribution of the EOO algorithm is given as follows.

1) The precision of selecting a mussel by calculating the time needed to break one is calculated using the bird's energy and the mussel's size as variables to estimate the anticipated location of the desired food.
2) The random numbers entered during optimization help investigate new areas during each cycle. Avoid a local minimum issue as a result.
3) The random numbers used at each optimization stage ensure research and application.

| Algorithm 1: Suggested MML-EOO pseudocode |
|---|
| Begin with assigning the population of the EO $P_r = (r = 1,2,..,s)$ |
| Determine the fitness of all the EO |
| Determine the best EO $P_m$ |
|   For$(r = 1 to R)$ |
|   For$(p = 1 to P)$ |
|     While$(r > 0)$ |
|       For every ideal oyster |
|         **Determine the size of the oyster $O$ using the fitness-based concept using Eq. (6)** |
|         Compute the time required to open the oyster $J$ using Eq. (9) |
|         Compute the energy that is currently available in the EO $D$ using Eq. (10) |
|         Compute the caloric value gained by consuming the oyster $Q$ using Eq. (11) |
|         Utilizing Eq. (7) and Eq. (8) upgrade the position of the oyster |
|       End for |
|       Determine the fitness of all search agent |
|       Amend the best EO |
|     End while |
|     Return $P_m$ |
|   End for |
|   End for |
| End |

The flowchart of the suggested MML-EOO algorithm is given in Appendix (A) supplementary information. The pseudocode of the proposed MML-EOO algorithm is provided below in Algorithm 1.

## 6. Results and Discussion

### 6.1 Experimental Setup

The constructed DL-based BC detection model was assessed using the Python platform. The experimental results of this evaluation were further discussed. The DL-based BC detection model was constructed with an iteration count that should not exceed 50 and a maximum population size of 10, respectively. The MML-EOO-ACA-ATRUNet-AMDN-based BC detection framework was assessed against different classifiers like, UNet [32], KNN [29], CNN [25], XGBoost [24], ResUNet [33], SwinUNet [34], and Deeplab [35], and contrasted with existing meta-heuristic algorithms like Grey Wolf Optimization algorithm (GWO)-ACA-ATRUNet-AMDN [36], Honey Badger Algorithm (HBA)-ACA-ATRUNet-AMDN [37], JAYA-ACA-ATRUNet-AMDN [38], and EOO-ACA-ATRUNet-AMDN [39] algorithm for representing the accuracy of the developed deep learning-based BC detection model.

### 6.2 Validation Metrics Used in Evaluation

The below-provided metrics are utilized in assessing the implemented BC detection framework.

| $Np = \frac{VW}{TV+VW}$ (12) | $pcn = \frac{TU}{TU+TV}$ (13) | $Fd = \frac{TV}{TV+TU}$ (14) |
|---|---|---|
| $Sp = \frac{VW}{VX+TU}$ (15) | $Se = \frac{TU}{TU+VX}$ (16) | $Fs = \frac{2*TU}{2*(TU+TV+VX)}$ (17) |

| $Fn = \frac{VX}{TU+VX}$ (18) | $Jd(MI_{ma}^{am}, SI_{ad}^{TRU}) = \frac{|MI_{ma}^{am} \cap SI_{ad}^{TRU}|}{|MI_{ma}^{am} \cup SI_{ad}^{TRU}|}$ (19) |
|---|---|
| $Mc = \frac{TU*VW - TV*VX}{\sqrt{(TU+TV)(TU+VX)(VW+TV)(VW+VX)}}$ (20) | |

Negative Predictive Value (NPV) $Np$ is determined by Eq. (12). The precision $pcn$ is evaluated based on Eq. (13). False Discovery Rate (FDR) $Fd$ is computed as in Eq. (14). Specificity $Sp$ is determined as in Eq. (15). Matthews Correlation Co-efficient (MCC) $Mc$ is evaluated as provided in Eq. (16). Sensitivity $Se$ is calculated using Eq. (17). The F1 score Fs is determined using Eq. (18). False Negative Rate (FNR) $Fn$ is evaluated using Eq. (19). The Jaccard distance $Jd$ between the ground truth image/mask images and the segmented image is computed using Eq. (20).

## 6.3 Experimental Outcome

The segmented BC mammogram images obtained from various deep learning technique and the ground truth comparison with the suggested MML-EOO-ACA-ATRUNet technique output is shown in Figure. 4.

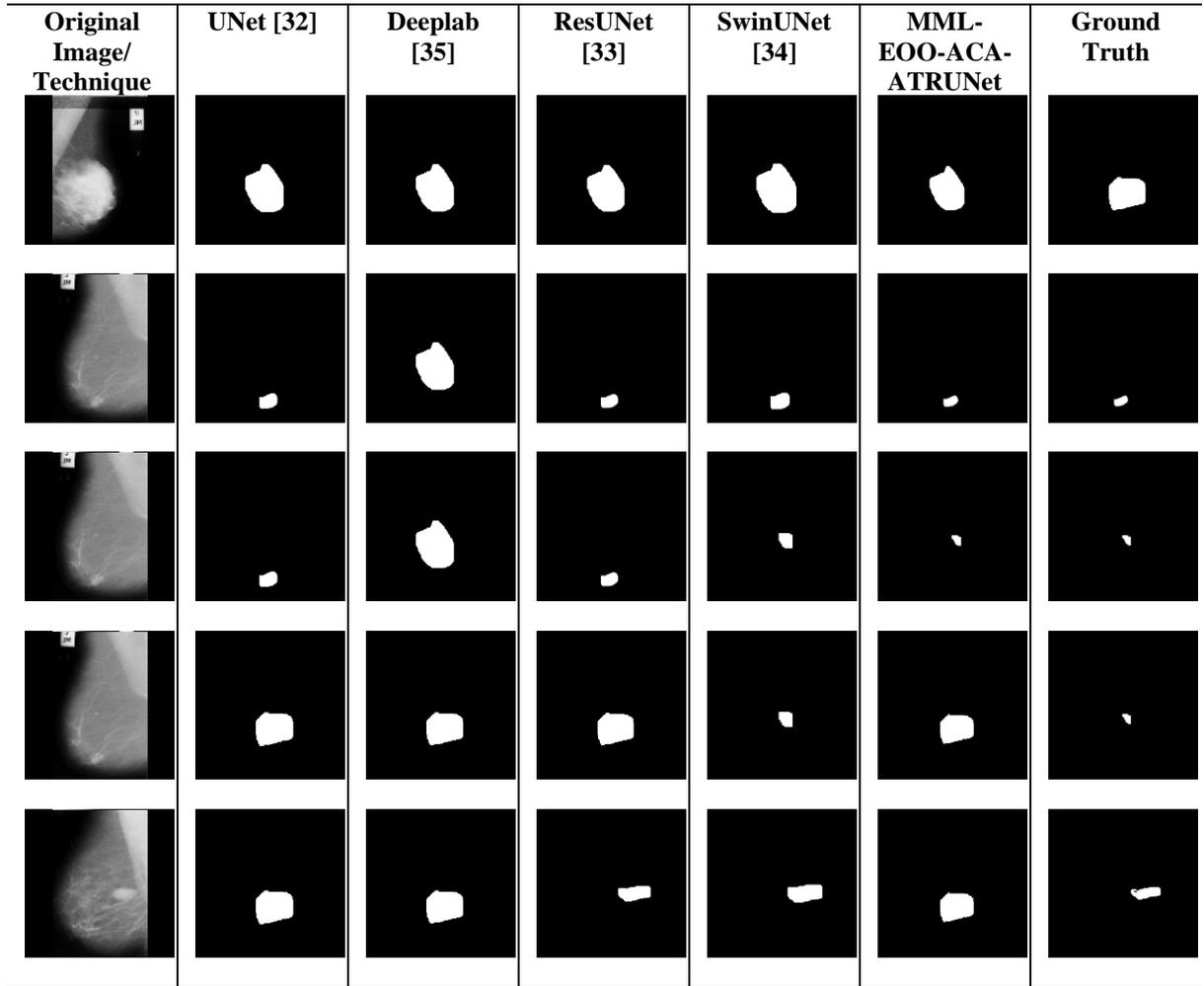

**Figure 4.** Segmented mammogram BC image outputs from proposed and conventional classifiers.

## 6.4 Performance Comparison of the Developed BC Detection Model with Conventional Classifiers

The performance comparison of the developed MML-EOO-ACA-ATRUNet-MDN BC detection model with respect to conventional classifiers for Dataset 1 and Dataset 2 is given in Figure. 5, and Figure. 6, respectively. The precision of the implemented MML-EOO-ACA-ATRUNet-MDN 5%, 2.56%, 3.8%, and 5.56% higher than the KNN, CNN, RAN, and RAN-LSTM classifiers for dataset 1 for ReLU activation function, respectively. We have mentioned the precision and accuracy results here; other evaluation measures' performance comparison results can be found in Appendix A: supplementary material.

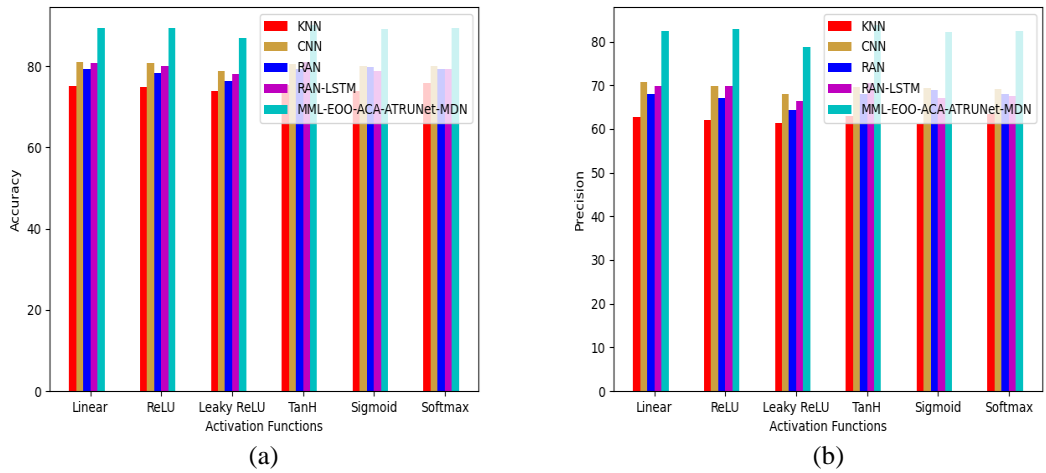

**Figure 5.** Performance comparison of the developed bc detection model with conventional classifiers with respect to dataset 1 in terms of "(a) accuracy, (b) precision."

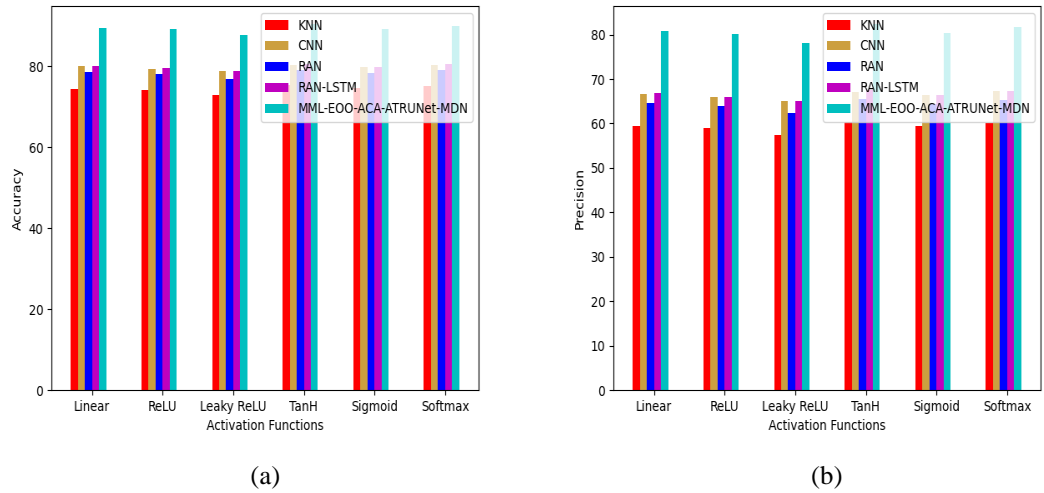

**Figure 6.** Performance comparison of the developed bc detection model with conventional classifiers with respect to dataset 2 in terms of "(a) accuracy, (b) precision."

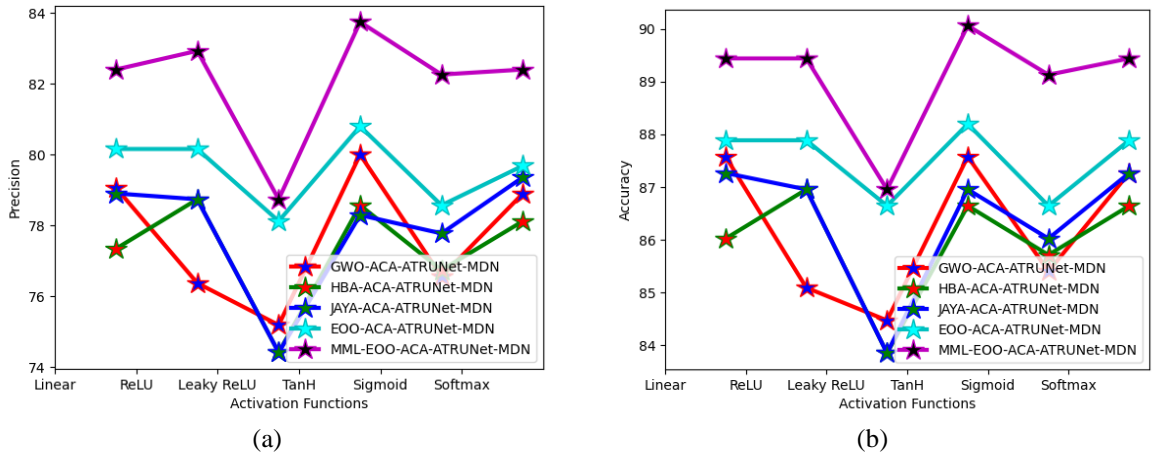

**Figure 7.** Evaluation of the recommended bc detection framework with existing algorithms with respect to dataset 1 in terms of "(a) precision (b) accuracy."

## 6.5 Comparison of the Proposed BC Detection Framework with Existing Algorithms

The performance comparison of the proposed MML-EOO-ACA-ATRUNet-MDN BC detection model with respect to various existing algorithms for Dataset 1 and Dataset 2 is given in Figure. 7 and Figure. 8, respectively. The accuracy of the proposed MML-EOO-ACA-ATRUNet-MDN-based BC detection framework is 2.32%, 3.27%, 3.39%, and 3.63% better than the EOO-ACA-ATRUNet-MDN, JAYA--ACA-ATRUNet-MDN, HBA-ACA-ATRUNet-MDN, and GWO-ACA-ATRUNet-MDN algorithms respectively for dataset 2 on Leaky ReLU activation function.

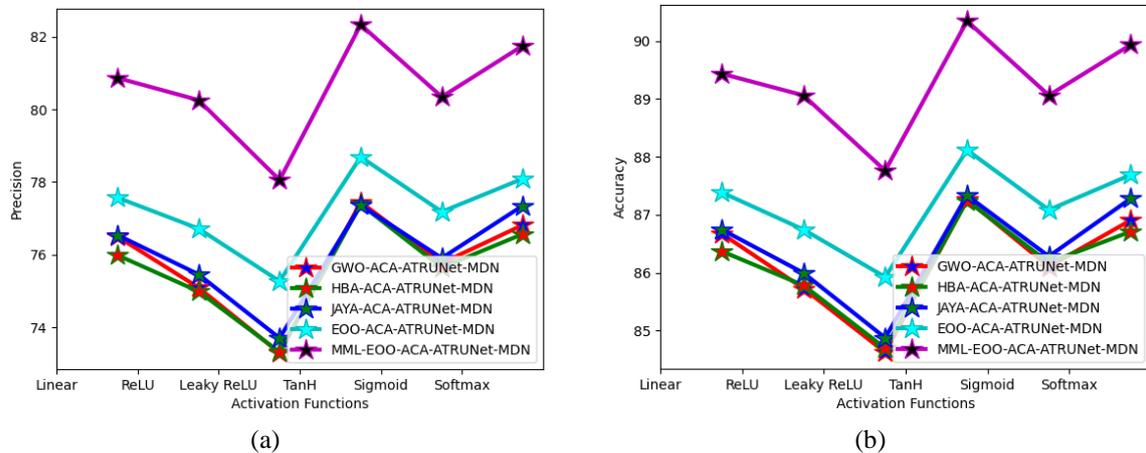

(a)          (b)

**Figure 8.** Evaluation of the recommended bc detection framework with existing algorithms with respect to dataset 2 in terms of "(a) precision, (b) accuracy".

## 6.6 Statistical Examination of the Implemented BC Detection Framework with Traditional Classifiers

The statistical examination of the implemented MML-EOO-ACA-ATRUNet-MDN-based BC detection framework to different traditional classifiers in Dataset 1 and Dataset 2 is shown in Figure. 9 and Figure. 10, respectively. The precision of the implemented MML-EOO-ACA-ATRUNet-MDN BC detection framework is 2.63%, 1.33%, 5.13%, and 5.71% best than the SwinUNet, Deeplab, ResUNet, and UNet classifiers respectively for dataset 1.

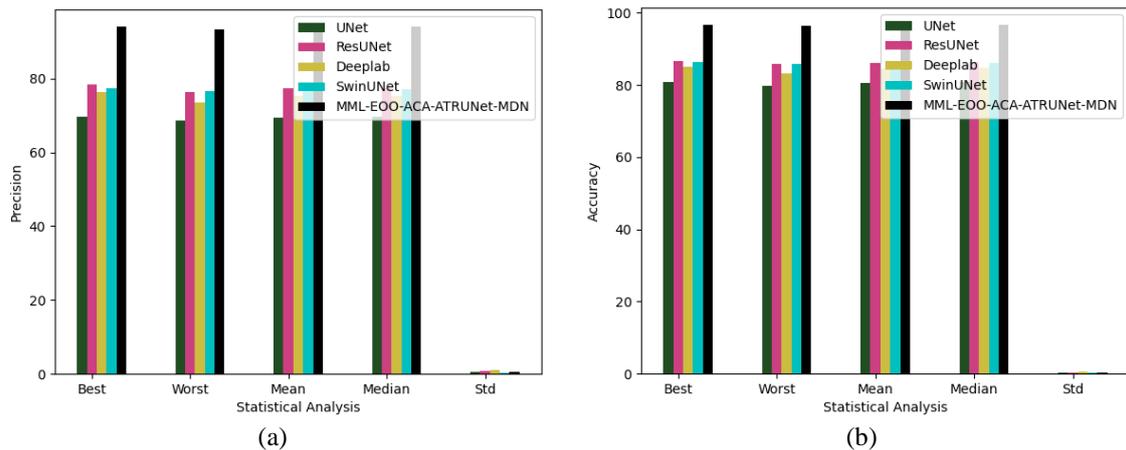

(a)          (b)

**Figure 9.** Statistical examination of the implemented bc detection framework with traditional classifiers with respect to dataset 1 in terms of "(a) precision, (b) accuracy."

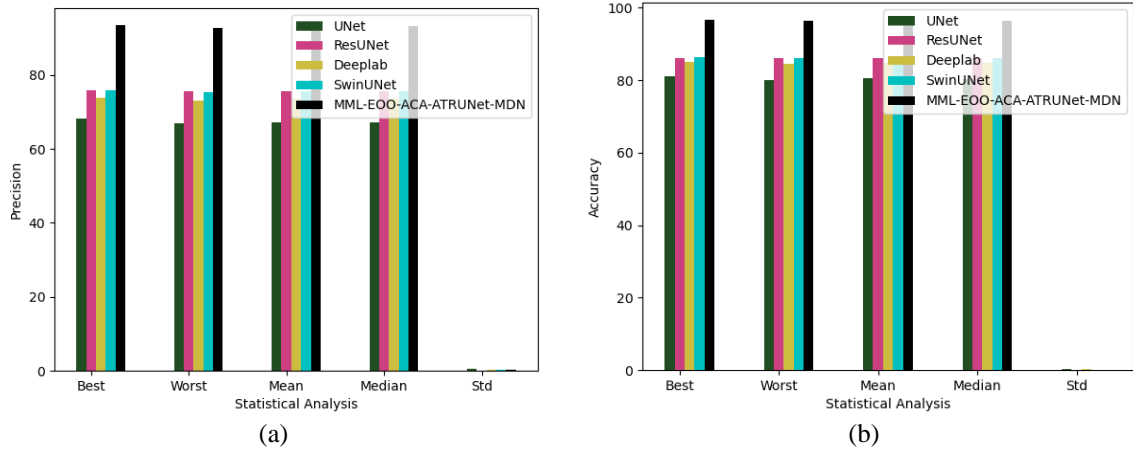

(a)                                            (b)

**Figure 10.** Statistical examination of the implemented bc detection framework with traditional classifiers with respect to dataset 2 in terms of "(a) precision, (b) accuracy."

### 6.5 Statistical Assessment of the Constructed BC Detection Model with Other Heuristic Algorithms

The statistical assessment of the constructed MML-EOO-ACA-ATRUNet-MDN BC detection model to other heuristic algorithms in Dataset 1 and Dataset 2 is illustrated in Figure. 11 and Figure. 12, respectively. The sensitivity of the constructed MML-EOO-ACA-ATRUNet-MDN BC detection model is 1.04%, 2.11%, 2.65%, and 2.11% higher than the EOO-ACA-ATRUNet-MDN, JAYA-ACA-ATRUNet-MDN, HBA-ACA-ATRUNet-MDN, and GWO-ACA-ATRUNet-MDN algorithms respectively for dataset 1.

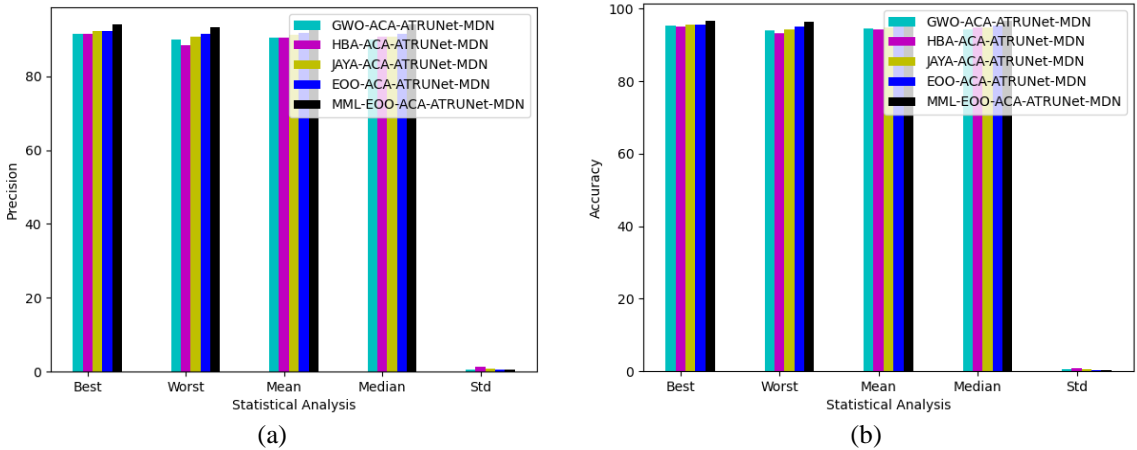

(a)                                            (b)

**Figure 11.** Statistical assessment of the constructed bc detection model with other heuristic algorithms with respect to dataset 1 in terms of "(a) precision, (b) accuracy."

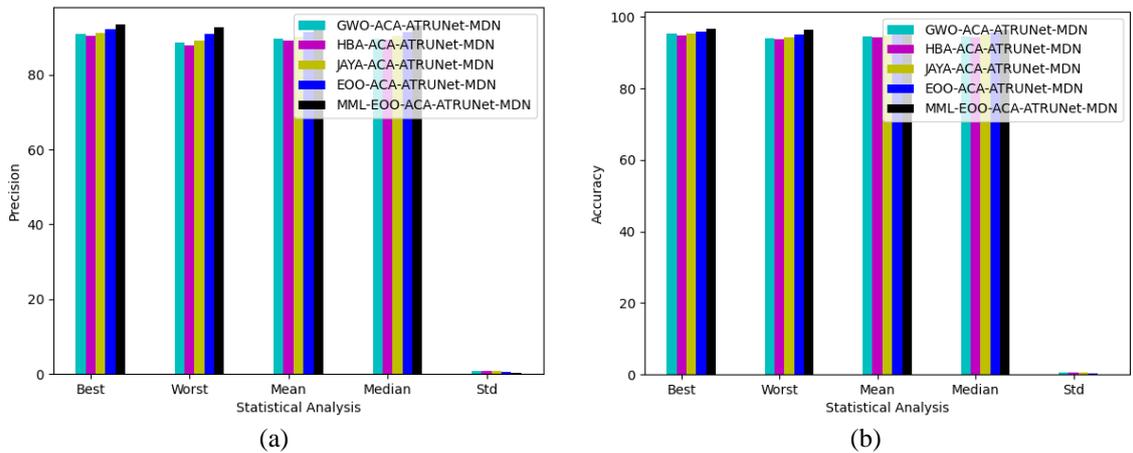

(a)                                            (b)

**Figure 12.** Statistical assessment of the constructed bc detection model with other heuristic algorithms with respect to dataset 2 in terms of "(a) precision, (b) accuracy."

### 6.8 Cost Function Analysis of the Generated BC Detection Framework

The convergence analysis of the generated MML-EOO-ACA-ATRUNet-MDN-based BC detection framework is depicted in Fig. 18. The cost function of the generated MML-EOO-ACA-ATRUNet-MDN-based BC detection framework is 7.87%, 9.4%, 13.11%, and 23.19% lesser than the JAYA-ACA-ATRUNet-MDN, GWO-ACA-ATRUNet-MDN, HBA-ACA-ATRUNet-MDN, and EOO-ACA-ATRUNet-MDN algorithms respectively for dataset 2.

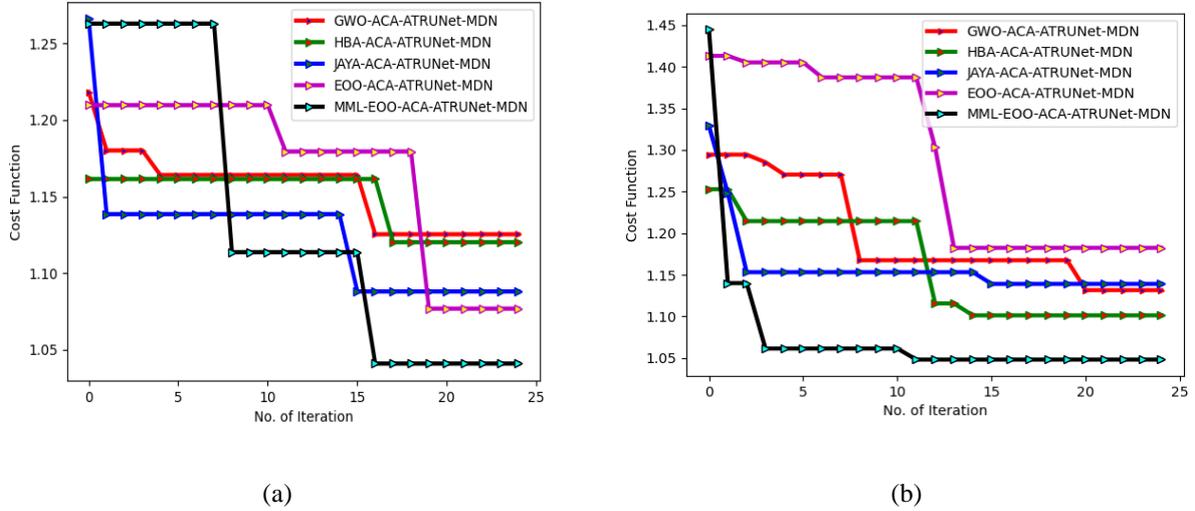

(a) (b)

**Figure 13.** Cost function analysis of the generated bc detection framework in terms of "(a) Dataset 1, and (b) Dataset 2."

### 6.9 Classifier Analysis of the Suggested MML-EOO-ACA-ATRUNet-MDN-based BC Detection Framework

The performance comparison of the suggested MML-EOO-ACA-ATRUNet-MDN-based BC detection framework is tabulated in Table 3. The precision of the suggested MML-EOO-ACA-ATRUNet-MDN-based BC detection framework is 34.52%, 18.41%, 19.32%, and 22.49% higher than the KNN, CNN, XGBoost, and ACA-ATRUNet-MDN classifiers respectively for dataset 1.

**Table 3.** Performance Analysis of The Suggested BC Detection Framework with Conventional Classifiers.

| TERMS/Classifiers | KNN [29] | CNN [25] | XGBoost [24] | ACA-ATRUNet-MDN [40] | **MML-EOO-ACA-ATRUNet-MDN** |
|---|---|---|---|---|---|
| Dataset 1 | | | | | |
| NPV | 83.607 | 87.435 | 87.368 | 87.568 | 93.434 |
| Accuracy | 73.913 | 80.124 | 79.814 | 78.882 | 89.130 |
| F1-Score | 66.929 | 73.984 | 73.684 | 73.016 | 85.356 |
| Specificity | 73.913 | 80.676 | 80.193 | 78.261 | 89.372 |
| Precision | 61.151 | 69.466 | 68.939 | 67.153 | 82.258 |
| FPR | 26.087 | 19.324 | 19.807 | 21.739 | 10.628 |
| MCC | 0.463 | 0.583 | 0.578 | 0.565 | 0.769 |
| Sensitivity | 73.913 | 79.130 | 79.130 | 80.000 | 88.696 |
| FDR | 38.849 | 30.534 | 31.061 | 32.847 | 17.742 |
| FNR | 26.087 | 20.870 | 20.870 | 20.000 | 11.304 |
| Dataset 2 | | | | | |
| FPR | 25.254 | 20.069 | 21.593 | 20.176 | 10.877 |
| Sensitivity | 74.185 | 79.583 | 78.087 | 79.904 | 88.936 |
| Accuracy | 74.559 | 79.815 | 78.300 | 79.850 | 89.061 |
| NPV | 85.274 | 88.675 | 87.739 | 88.820 | 94.156 |
| Specificity | 74.746 | 79.931 | 78.407 | 79.824 | 89.123 |
| Precision | 59.494 | 66.473 | 64.390 | 66.444 | 80.348 |
| F1-Score | 66.032 | 72.440 | 70.580 | 72.555 | 84.424 |
| FNR | 25.815 | 20.417 | 21.913 | 20.096 | 11.064 |
| MCC | 0.468 | 0.573 | 0.543 | 0.575 | 0.763 |
| FDR | 40.506 | 33.527 | 35.610 | 33.556 | 19.652 |

## 6.10 Algorithmic Evaluation of the Recommended MML-EOO-ACA-ATRUNet-MDN-based BC Detection Model

The algorithmic evaluation of the recommended MML-EOO-ACA-ATRUNet-MDN-based BC detection framework is tabulated in Table 4. The NPV of the recommended MML-EOO-ACA-ATRUNet-MDN-based BC detection model is 1.94%, 1.64%, 1.68%, and 1.2% better than the GWO-ACA-ATRUNet-MDN, HBA-ACA-ATRUNet-MDN, JAYA-ACA-ATRUNet-MDN, and EOO-ACA-ATRUNet-MDN algorithms respectively for dataset 2.

**Table 4.** Algorithmic Evaluation of The Recommended BC Detection Model

| TERMS/ Algorithm | GWO-ACA-ATRUNet-MDN [36] | HBA-ACA-ATRUNet-MDN [37] | JAYA-ACA-ATRUNet-MDN [38] | EOO-ACA-ATRUNet-MDN [39] | **MML-EOO-ACA-ATRUNet-MDN** |
|---|---|---|---|---|---|
| Dataset 1 | | | | | |
| FPR | 14.493 | 14.493 | 13.527 | 13.043 | 10.628 |
| FDR | 23.438 | 23.256 | 22.222 | 21.429 | 17.742 |
| Sensitivity | 85.217 | 86.087 | 85.217 | 86.087 | 88.696 |
| NPV | 91.237 | 91.710 | 91.327 | 91.837 | 93.434 |
| Precision | 76.563 | 76.744 | 77.778 | 78.571 | 82.258 |
| Accuracy | 85.404 | 85.714 | 86.025 | 86.646 | 89.130 |
| FNR | 14.783 | 13.913 | 14.783 | 13.913 | 11.304 |
| Specificity | 85.507 | 85.507 | 86.473 | 86.957 | 89.372 |
| F1-Score | 80.658 | 81.148 | 81.328 | 82.158 | 85.356 |
| MCC | 0.692 | 0.700 | 0.704 | 0.717 | 0.769 |
| Dataset 2 | | | | | |
| Specificity | 86.291 | 86.104 | 86.317 | 87.146 | 89.123 |
| NPV | 92.363 | 92.639 | 92.603 | 93.039 | 94.156 |
| MCC | 0.700 | 0.703 | 0.705 | 0.721 | 0.763 |
| Precision | 75.768 | 75.644 | 75.906 | 77.182 | 80.348 |
| FPR | 13.709 | 13.896 | 13.683 | 12.854 | 10.877 |
| Accuracy | 86.104 | 86.175 | 86.282 | 87.084 | 89.061 |
| FDR | 24.232 | 24.356 | 24.094 | 22.818 | 19.652 |
| F1-Score | 80.441 | 80.629 | 80.731 | 81.779 | 84.424 |
| FNR | 14.270 | 13.683 | 13.789 | 13.041 | 11.064 |
| Sensitivity | 85.730 | 86.317 | 86.211 | 86.959 | 88.936 |

## 7. Conclusion

An accurate deep learning-based BC detection framework has been constructed and evaluated successfully. The mammography images were initially gathered from benchmark image sources of mammogram data. After that, the segmentation process took place, which was carried out by developing ACA-ATRUNet, which was built by combining the Transformer block, ResNet, and UNet. The parameters in the implemented ACA-ATRUNet were tuned and adjusted with the aid of the modified MML-EOO algorithm. Finally, ACA-AMDN was implemented to complete the detection process. The same MML-EOO was adopted in tuning the parameters of the ACA-AMDN. The constructed MML-EOO-ACA-ATRUNet-AMDN model's BC detection performance was verified and assessed using several measures and contrasted with conventional methods. From the assessment, it was proved that the precision of the suggested MML-EOO-ACA-ATRUNet-MDN-based BC detection framework is 34.52%, 18.41%, 19.32%, and 22.49% higher than the KNN, CNN, XGBoost, and ACA-ATRUNet-MDN classifiers respectively. Hence, the suggested MML-EOO-ACA-ATRUNet-MDN-based BC detection framework could effectively and accurately detect BC in women with any issues with maximum precision and minimum error.

## 8. Appendix A: Supplementary Material

**Detailed View of Atrous Convolution-based Attentive and Adaptive Breast Cancer Segmentation Model using Mammogram Images**

**UNet**

The level of UNet's contribution has increased for the segmentation of medical images. The fine-tuned approach is frequently used in radiology assessment to swap the Fully Connected Layers (FCL) of the pre-trained framework with a new range of FCL for retraining an original dataset while fine-tuning the kernels in the pre-trained convolution through the backpropagation technique. The segmentation technique in medical imaging entails categorising every pixel into potential components. One of the most commonly used systems for this job is UNet [26], designed to consider medical imaging alone. This method is designed based on Fully Connected Network (FCN) architecture, with convolution layers and no dense layers, allowing it to accept images of every resolution. Convolutional and clustering layers combine to generate encoder-decoder architecture in the UNet architecture. The design is mostly made up of synthesising and evaluation channels. The contraction route remains the first channel that constitutes the network's encoder part. This takes the structure of CNN as a reference. An up-sampling layer is usually preceded by a deconvolution layer in the synthesis operation, known as the expansion stage or decoder. Just on the apex of the omni-channel feature maps, the U-shaped symmetric structure is depicted as a box with a varied channel count. The diagram's arrows indicate various operations; for example, a black arrow in every step represents two $1 \times 1$ convolutional procedures. With $2 \times 2$ the size and a step of 2, the max-pooling procedure, which reduces the image's resolution in half, is indicated by the purple arrow. During each down-sampling step, the procedure is repeated four times, with the number of the convolution layer's filters being multiplied. Finally, a succession of two convolutional procedures connects the encoder to the decoder. The feature maps are up-sampled using $2 \times 2$ inverted convolutional processes in the decoder. The two convolutions and up-sampling processes are carried out four times repeatedly, similar to that in the encoder, and each step reduces the total amount of filters to half. The blue arrow indicates where the image is resized or increased to its initial dimensions. The source image's size is set to $128 \times 128 \times 3$. The left portion shows each step's two successive max-pooling and convolution layers. The term $F_1, F_2, \ldots, F_4$ denotes the direction of output from the max-pooling layer, $B_1, B_2, \ldots, B_9$ represents the direction of output from the convolution layer, and $E_6, E_7, \ldots, E_9$ indicates the direction of output from the upsampling layer. To obtain the final segmented result, $a\ 1 \times 1$ convolution procedure is performed on the right-hand side of the design. The features of the low-level layer are merged with high-level layers to maintain the portion of the spatial information. The decoder conducts up-sampling once the encoder obtains the characteristics. In the dimension of channels, the UNet uses splicing and fusing. The network may also be trained with less input because its structure converges quickly. UNet can segment images quite effectively due to its design. After raining the model, every convolutional layer in the networks utilises filters to retrieve the ideal traits. The convolutional layers are governed by the underlying mathematical formula in Eq. (21).

$$G_a^{(c)} = d^{(c)}\left(\sum_{g=1}^{H} G_a^{(c-1)} * I_{ag}^{(c)} + e_a^{(c)}\right) \tag{21}$$

In Eq. (1), the term $G_a^{(c-1)}$ indicates the prior layer's output which is fed as input to the $c^{th}$ layer, $G_a^{(c)}$ indicates the output of the $c^{th}$ layer, $H$ indicates the prior layer's feature map counts, $e_a^{(c)}$ indicates the bias, $d$ indicates the activation function, and $I_{ag}^{(c)}$ indicates the filter at the $a^{th}$ layer. The diagrammatic representation of the UNet architecture is shown in Figure. 14.

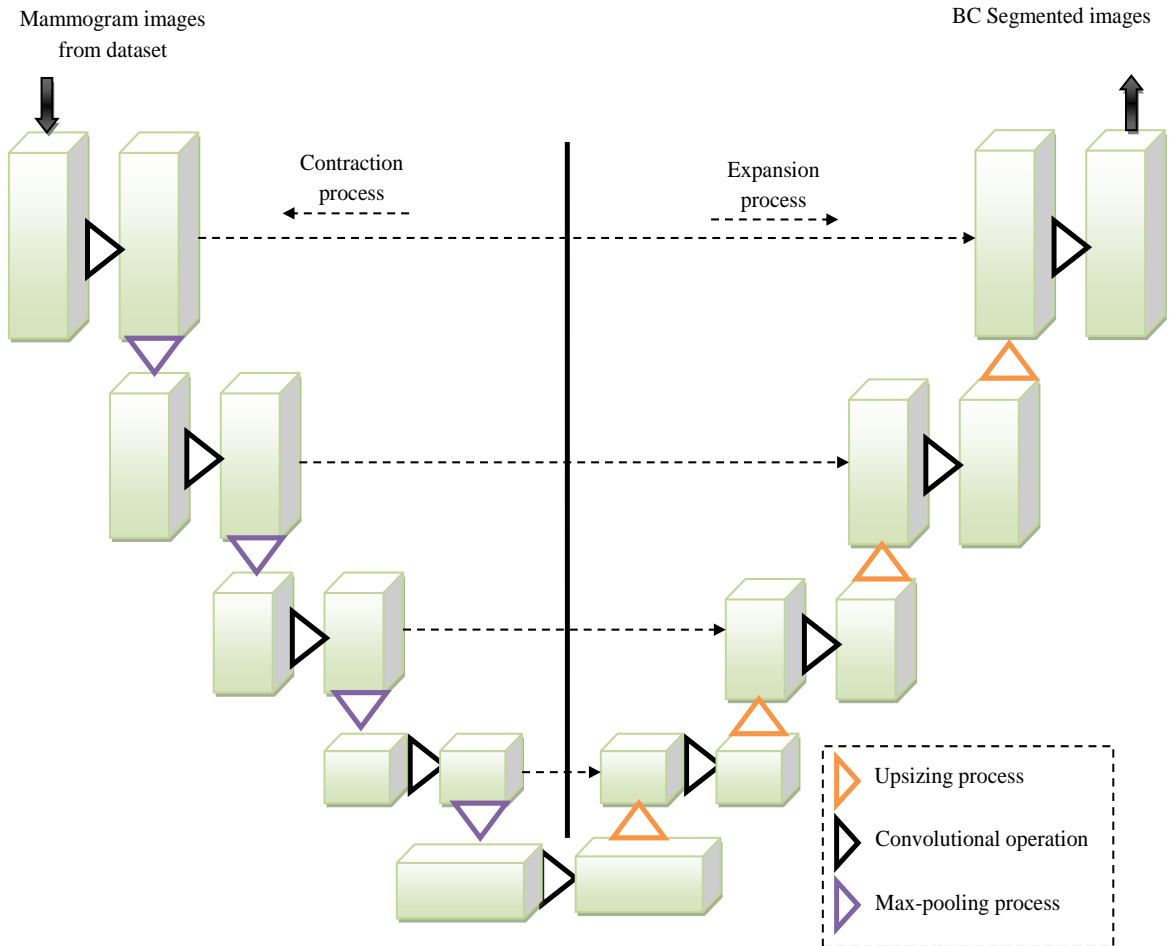

**Figure 14.** Diagrammatic representation of UNet architecture

**Trans-Res-UNet**

Combining the transformer block fused with the ResUNet forms the Trans-Res-UNet [27] structure. CNN is a widely adapted and popular framework in artificial intelligence. After properly extracting multi-dimensional features using the convolution kernels, it works with the pooling layer to minimise the input of every layer. Although deep networks frequently surpass them in classification, CNN is challenging to train for two key reasons: The first is the disappearing gradients: due to activation function, it is possible for a neuron to pass away during training and never reappear. Initiation approaches that attempt to start the optimisation process with an active group of neurons can solve this issue. The second is that the optimisation is more complex since training the network as the model adds more parameters becomes challenging. Sometimes, this is not a compliance issue because adding another layer might result in more training mistakes.

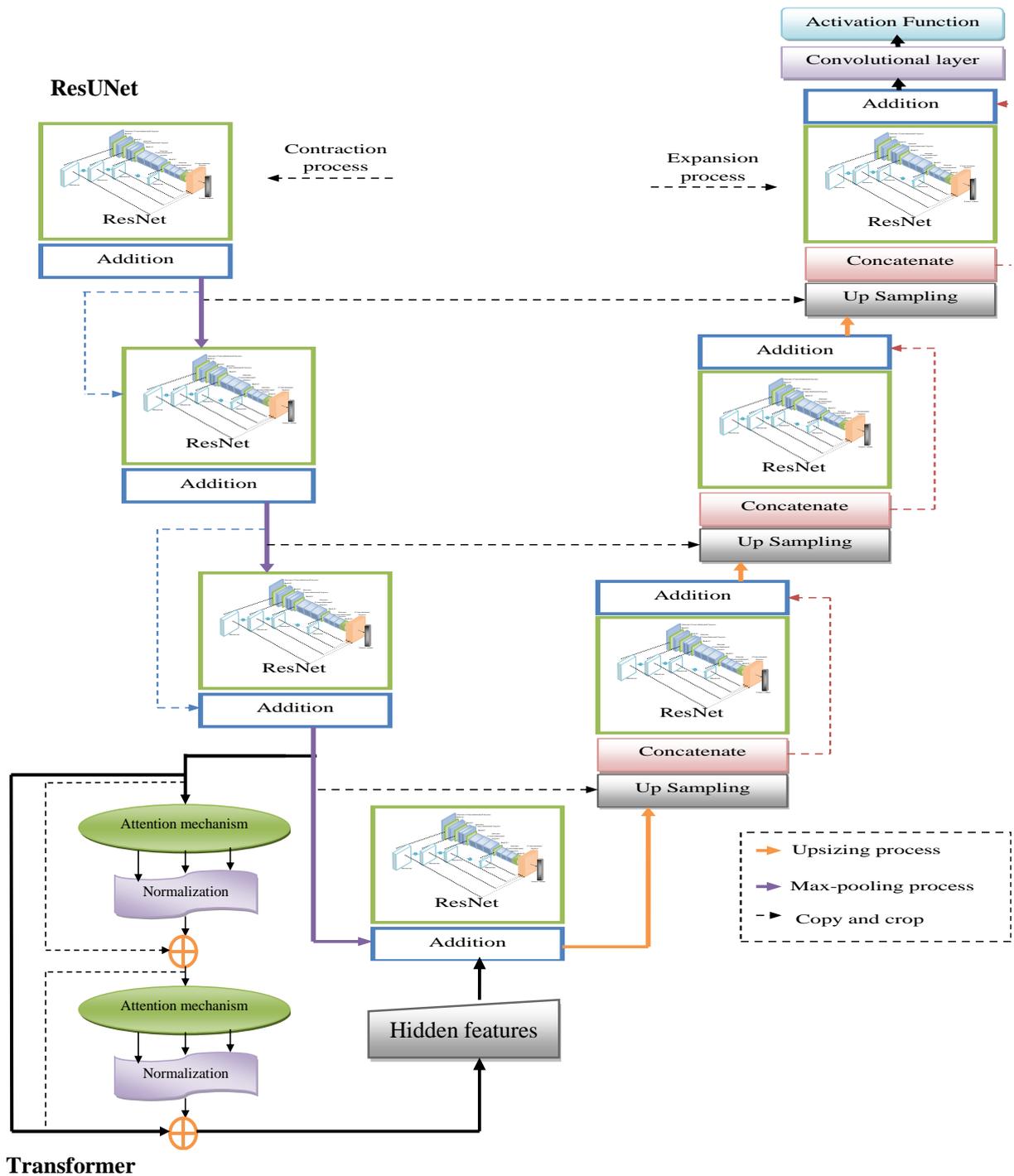

**Figure 15.** The architecture of the Trans-Res-UNet framework

Deep CNNs have greater classification performance but are more challenging to train. ResNet [28] is a useful tool to address these proposed issues. The key distinction between ResNet and conventional convolutional layers is that ResNet has shortcut links running parallel to them. Unlike the convolution layers, these shortcut linkages are always active, and gradients can readily propagate back, leading to faster training. On the one hand, introducing data shortcuts on CNN can aid in network convergence. CNN keeps its convolutional layer to extract features effectively. The convolutional layers are separated into numerous residual blocks when ResNet introduces the shortcuts. ResNet can be characterised as a sequence of multiple fundamental blocks, each with parallel shortcut links that can be added to its output. The shortcut connection is only a distinctive matrix if the basic block's input and output sizes are equal. If not, the size can be adjusted using zero padding (for increasing) and average pooling (for decreasing). After the addition module, adding a non-parameterised layer like Rectified Linear Unit (ReLU) offers neither a significant benefit nor a significant drawback. It has been found that overfitting causes the error to increase after a certain number of steps as the depth grows in classical CNN. To avoid this, a novel strategy known as "Residual Block" has been added to the ResNet model. The primary characteristic that sets it apart from other designs is how residual values are created by adding blocks and then fed into the model's subsequent levels. The flow of information in the residual block is given by Eq. (22).

$$A_{b-2} = C\left(D\left(L(A_{b-3})\right)\right) \tag{22}$$

$$A_b = A_{b-3} + A_{b-1} \tag{23}$$

In Eq. (22) and Eq. (23), the terms $L$ denote the 2D convolutional layer, $C$ denotes the batch normalisation layer, and $D$ denotes the ReLU activation layer. Convolutional layers are primarily responsible for feature extraction of the state in forward propagation. Yet, in backpropagation, the loss signal travels through the more active shortcuts to the shallow layer. Eventually, ResNet architecture can secure the establishment of a successful network and raise the success rate, which assures confidential decision-making. The task of change detection was viewed as an extension of the image segmentation task. Researchers widely used the UNet network, and multiple studies have shown that this network delivered accurate segmentation findings. UNet employed jumping connections to integrate high-level semantic data and low-level traits to achieve feature extraction and detail recovery produced by Tran-ResUNet. CNN did pretty well but had problems acquiring reliable global information because convolutional processes are local. Transformer, in comparison, gathered global contextual information using the self-attention method. This made it possible for Transformer to accurately model spatiotemporal global semantic linkages and make feature representation of interest shifts easier. Transformers can be used in conjunction with UNet and ResNet to combine the strengths of each technology and enhance image segmentation results. Trans-Res-UNet is a hybrid model combining CNN and Transformer that can handle both local and international news at once. The segmentation result was acquired by upsampling and skip connection step by step after the UNet decoder after the feature map had been retrieved using CNN and then input into the Transformer encoder module. It was possible to gather enough data by fusing the high-resolution feature map with the up-sampled feature map via skip connections. This model has been used to segment medical images with great effectiveness. The architecture of the Trans-Res-UNet is given in Figure. 15.

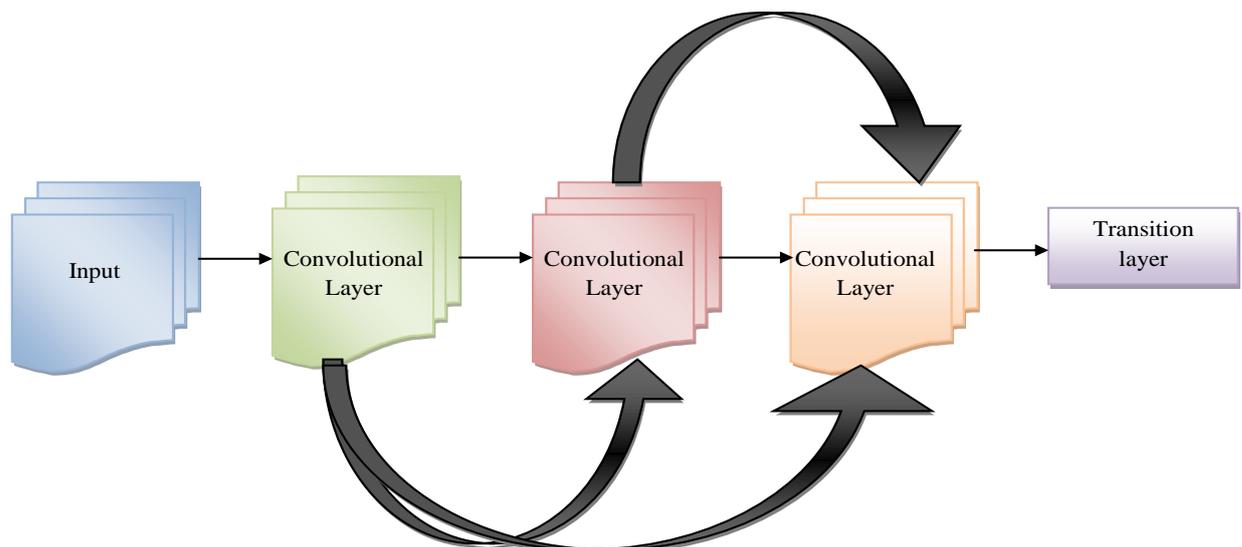

**Figure 16.** Illustration of multiscale densenet structure.

**Architectural Representation of Atrous Convolution-based Attentive and Adaptive Breast Cancer Detection Model using Mammogram Images**

**Multi-scale DenseNet**

A DenseNet [29] is a type of CNN that uses the dense connections made by the dense block between the layers. DenseNet helps increase accuracy for high-level neural networks affected by disappearing gradients. Because of the longer, denser paths between the output and input layers, the pictures disappear before they reach their destination. To reuse the feature maps, get rid of layer dependencies, and reduce the vanishing gradient issue, DenseNet is used. The DenseNet is used to create deeper learning networks. It is composed of transitional layers that are sandwiched between several dense blocks. The primary component module of DenseNet, dense block, connects all network levels directly while maintaining maximal data flow between layers. The $i^{th}$ layer output is computed as given in Eq. (24).

$$Q_i = M_i(q_0, q_1, \ldots, q_{i-1}) \tag{24}$$

The term $q_i$ in Eq. (7) denotes the current output $i-1$ denotes the former layer and $M_i(\bullet)$ represents the transformation of the sequence. The Multiscale DenseNet is nothing but repeating the process more than two times. The illustration of Multiscale DenseNet is shown in Figure. 16.

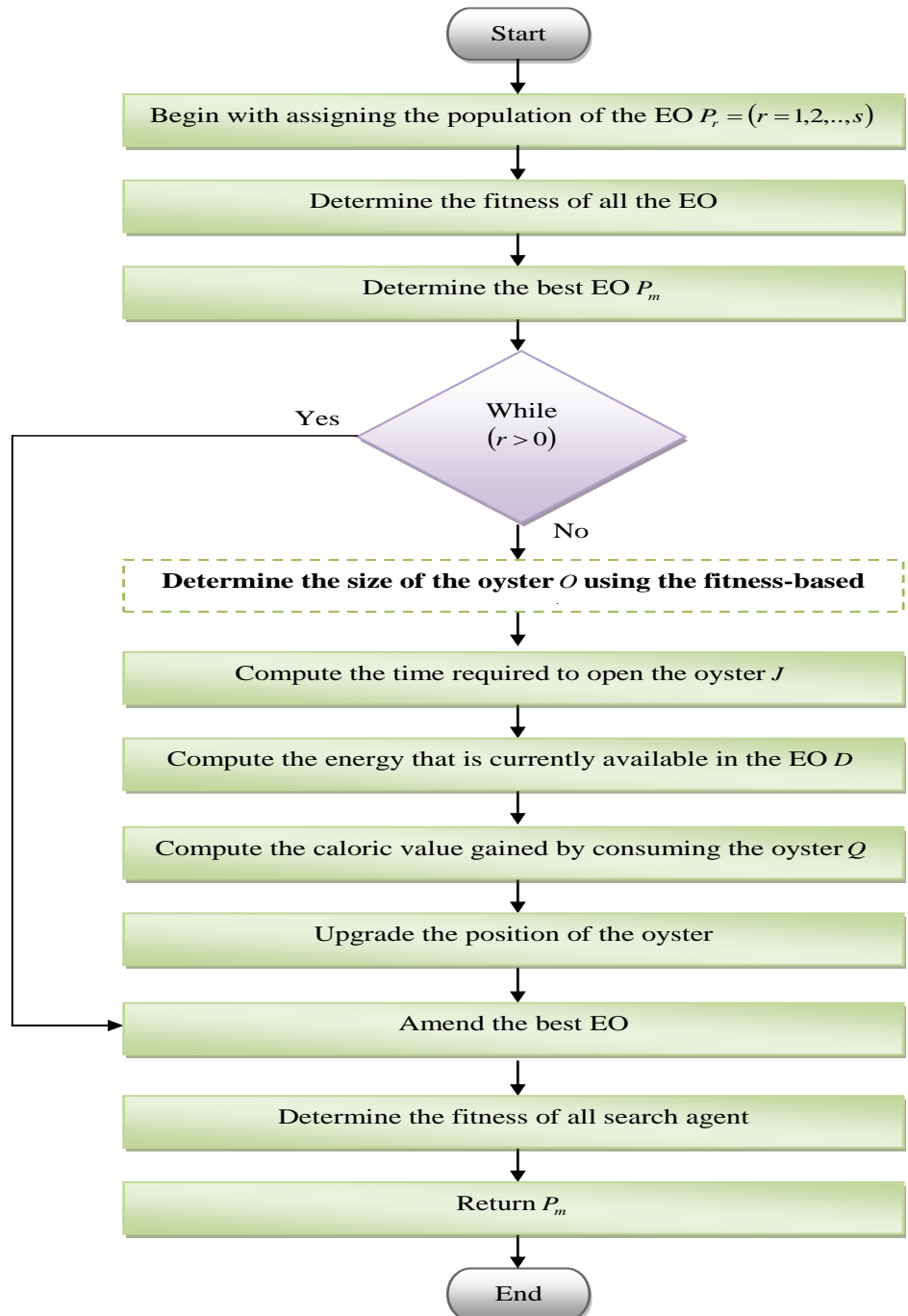

**Figure 17.** Flowchart of the recommended MML-EOO algorithm.